\documentclass[a4paper,11pt]{article}
\usepackage{pos}
\usepackage{wrapfig}

\newcommand{\chandra}{{\it Chandra}}

\newcommand{\suzaku}{{\it Suzaku}}
\newcommand{\xmm}{{\it XMM-Newton}}

\newcommand{\nustar}{\textit{NuSTAR}}
\newcommand{\fermi}{\textit{Fermi}}

\newcommand{\lumcgs}{ergs~s$^{-1}$}

\def\amin{\ifmmode^{\prime}\else$^{\prime}$\fi}
\def\asec{\ifmmode^{\prime\prime}\else$^{\prime\prime}$\fi}
\def\simgt{\lower.5ex\hbox{$\; \buildrel > \over \sim \;$}}
\def\simlt{\lower.5ex\hbox{$\; \buildrel < \over \sim \;$}}

\def\edot{$\dot E$}

\newcommand{\apj}{ApJ}
\newcommand{\apjs}{ApJS}
\newcommand{\apjl}{ApJL}
\newcommand{\nat}{Nature}
\newcommand{\aap}{A\&A}
\newcommand{\mnras}{MNRAS}
\newcommand{\ssr}{Space Science Reviews} 

\title{NuSTAR broad-band X-ray observational campaign of energetic pulsar wind nebulae in synergy with VERITAS, HAWC and Fermi gamma-ray telescopes}
 \ShortTitle{NuSTAR X-ray observations of energetic pulsar wind nebulae}

\author*[a]{Kaya Mori}
\author[b]{Hongjun An} 
\author[a]{Daniel Burgess}
\author[c]{Massimo Capasso}
\author[d]{Brenda Dingus}
\author[e]{Joseph Gelfand} 
\author[a]{Charles Hailey} 
\author[a]{Brian Humensky}  
\author[d]{Kelly Malone} 
\author[c]{Reshmi Mukherjee} 
\author[f]{Nahee Park} 
\author[a]{Isaac Pope}
\author[g]{Stephen Reynolds}  
\author[h]{Samar Safi-Harb}
\author[a]{Jooyun Woo}
\author[]{Galactic TeV source collaboration} 

\affiliation[a]{Columbia Astrophysics Laboratory, 
  550 West 120th Street, New York, NY 10027, USA}

\affiliation[b]{Chungbuk National University, 
Chungdae-ro 1, Seowon-Gu, Cheongju, Chungbuk, 28644  South Korea}

\affiliation[c]{Department of Physics and Astronomy, Barnard college, 3009 Broadway, New York, NY 10027, USA} 

\affiliation[d]{Los Alamos National Laboratory, Los Alamos, NM 87545, USA}

\affiliation[e]{NYU Abu Dhabi, PO Box 129188, Abu Dhabi, United Arab Emirates} 

\affiliation[f]{Department of Physics, Engineering Physics and Astronomy, 64 Bader Lane, 
Queen's University, Kingston, ON
Canada, K7L 3N6}

\affiliation[g]{Department of Physics, North Carolina State University, Raleigh, NC 27695-8202, USA} 

\affiliation[h]{Department of Physics and Astronomy,  University of Manitoba, 
Winnipeg, MB R3T 2N2, Canada}

\emailAdd{kaya@astro.columbia.edu}

\abstract{We report recent progress on the on-going \nustar\ observational campaign of 8  TeV-detected pulsar wind nebulae (PWNe). This campaign constitutes a major part of our \nustar\ study of some of the most energetic TeV sources in our Galaxy detected by VERITAS and HAWC. \nustar\ is the first focusing X-ray telescope operating above 10 keV in space with sub-arcminute angular resolution. Broad-band X-ray imaging and spectroscopy data, obtained by \nustar, allow us to probe sub-PeV electron populations through detecting synchrotron X-ray radiation. Our targets include PeVatron candidates detected by HAWC, the Boomerang nebula, PWNe crushed by supernova remnant shocks (or relic PWNe) and G0.9$+$0.1 in the Galactic Center. Using \fermi-LAT data and available TeV data, we aim to provide a complete, multi-wavelength view of a diverse class of middle-aged ($\sim$10--100 kyrs old) PWNe. Our \nustar\  analysis detected hard X-ray emission from the Eel and Boomerang PWNe and characterized their broad-band X-ray spectra most accurately. We plan to apply both time-evolution and multi-zone PWN models to multi-wavelength spectral energy distribution (SED) data over the radio, X-ray, GeV and TeV bands. In this proceeding, we will review our observational campaign and discuss the preliminary results for some PWNe.}

\FullConference{37$^{\rm{th}}$ International Cosmic Ray Conference (ICRC 2021)\\
		July 12th -- 23rd, 2021\\
		Online -- Berlin, Germany}

\begin{document}
\maketitle

\section{Introduction}

Relativistic leptons in the GeV to TeV energy band are ubiquitous throughout the Galaxy. 
It is generally believed that a majority of the energetic leptons are generated by pulsar wind nebulae (PWNe)  \citep{Cholis2018}. 
A PWN is an expanding bubble of a highly relativistic wind injected by a pulsar.  High energy observations detected synchrotron and Inverse Compton Scattering (ICS) emission from numerous PWNe in the X-ray and TeV bands, respectively, suggesting that non-thermal electrons are accelerated to GeV--PeV energies within the PWNe \citep{Kargaltsev2013}. 
The detailed characterization of PWNe's electromagnetic (EM)   spectra provides information on the particle's energy distribution produced by the pulsar, and also on the nature of acceleration in relativistic shocks. 
With their powerful shock-driven acceleration, PWNe can serve as the perfect laboratories to study the behavior of relativistic outflows.  But while the classic paper by \citet{KC1984} and subsequent work by \citet{Reynolds2009} laid the groundwork of particle acceleration and EM radiation from PWNe, other  factors, such as interactions with the SNR reverse shock and interstellar medium, can also play a role in affecting the evolution and properties of PWNe. 
As a result, the observed morphology and spectral properties of individual PWNe vary significantly. Thus, it is crucial to observe as many PWNe as possible at various ages and in different environments, in order to disentangle the fundamental PWN parameters, such as the electron energy distribution and magnetic field, that are responsible for driving the EM radiation from PWNe \citep{Torres2013}. Multi-wavelength spectral energy distributions (SEDs), coupled with spatially resolved spectroscopy (if possible), are required for this purpose.

Over the last two decades, the synergy between the X-ray and TeV gamma-ray telescopes has boosted our understanding of PWN astrophysics and electron acceleration in the Galaxy. From radio to MeV energy bands, synchrotron radiation dominates, while TeV emission originates from ICS of accelerated electrons off CMB photons or local IR/optical/UV radiation fields  \citep{Slane2017}. 
As of today, more than 70 X-ray emitting PWNe have been discovered and nearly half of them have been detected in the TeV band  \citep{Kargaltsev2013}. \chandra\ observations of PWNe revealed detailed X-ray features such as tori, jets and bow-shocks \citep{Kargaltsev2015}. On the other hand, \nustar, operating in the broad bandwidth from 3 to 79~keV, detected some of the highest energy electron populations ($E_e \simgt 100$~TeV) from PWNe \citep{Reynolds2017}. 
The earlier \nustar\ observations of nearby PWNe revealed the distinct hard  X-ray morphologies of several prominent sources, such as the Crab nebula \citep{Madsen2015}, G21.5$-$0.9 \citep{Nynka2014} and MSH~15$-$52 \citep{An2014}. 
Even when detailed spatial investigations are not feasible,  spectroscopic studies of unresolved or barely resolved PWNe in conjunction with other wavelength data can provide powerful constraints.  
Our \nustar\ observations of 8 PWNe, combined with multi-wavelength data, aim to address some fundamental astrophysics problems associated with TeV PWNe.

\smallskip 

\noindent (1) \textbf{Are some of the middle-aged PWNe PeVatrons?} Gamma-ray sources detected to the highest energies by HAWC \citep{Hawc2019} and LHAASO \citep{Cao2021} are of particular interest. These TeV sources may have a component powered by  relativistic hadrons interacting with molecular clouds  \citep{Bartko2008}. 
Alternatively, PeVatron PWNe that accelerate leptons to the PeV band have been proposed \citep{Arons2012} and some of the middle-aged ($\sim$10--100 kyrs old) PWNe may be  associated with gamma-ray sources detected above $\sim50$ TeV \citep{Abeysekara2020}. 
Along with the HAWC and LHAASO TeV telescopes which are sensitive above $\sim50$ TeV, \nustar\ plays a complementary role of  exploring sub-PeV electrons, with sub-arcminute angular resolution, via detecting the high-energy extension of their synchrotron radiation in the hard X-ray band \citep{Gelfand2019}. This was demonstrated by the \nustar\ detection of a spectral cut-off from 3C 58,  indicating that electrons are accelerated to $E_{\rm e} \sim 140$~TeV \citep{An2019}.  

\smallskip 
\noindent (2) \textbf{Does the electron energy distribution have multiple components and does it vary between PWNe?} 
Radio through X-ray spectra of PWNe require electron energy distributions $N(E_e) \propto E_e^{-p}$  (where $E_e$ is the electron energy) with the measured power-law index $p$ varying from as small as 1 for radio-emitting particles to as high as 2.8 for X-ray-emitting particles \citep{Reynolds2017}. The multi-wavelength SEDs of PWNe often fit to a broken power-law model \citep{Takata2011}. It is  unclear whether a single acceleration mechanism plus evolutionary effects can explain this,  or whether multiple independent electron components are required. 
 Broad-band SEDs can probe whether the universal electron acceleration mechanism (with similar $p$ values) operates for many PWNe or if other factors can later alter the electron energy distribution \citep{Kim2019}. However, only 9 bright/young PWNe yielded an accurate measurement of the X-ray emitting electron spectral index so far and they seem to vary from $p=1.6$ (Vela) to $p=2.8$ (Kes 75) \citep{Kargaltsev2015}. 

\smallskip 

\noindent (3) \textbf{What are the controlling parameters for distinct EM emission from middle-aged PWNe? } It is poorly understood which PWN parameters (e.g, pulsar age, B-field) primarily determine the various properties observed among middle-aged PWNe \citep{Kargaltsev2013, Kargaltsev2015}, which represent an  evolution phase from younger, well-confined PWNe to relic PWNe. Their complex, multi-wavelength features are also apparent from bow-shock PWNe \citep{Kargaltsev2017} and PWNe crushed by SN reverse shock through their interaction with the environment. 
There are also several middle-aged PWNe whose high-energy emission is not explained by a leptonic model alone. For example, \nustar\ observation of the evolved PWN DA 495 found that a hybrid SED model composed of both leptonic and hadronic components is required to account for the multi-wavelength SED \citep{Coerver2019}. 

\section{Observations and data analysis}

Our 8 targets for \nustar\ observations, listed in Table \ref{tab:nustar_targets}, represent a heterogeneous group of TeV PWNe in the pulsar age range of 7--43  kyrs with  spin-down power \edot $\sim 3\times10^{36}$ to $4\times10^{37}$~\lumcgs.  \nustar\ observed the 8 PWNe with a total exposure time of 590~ksec. 

\begin{table}[!ht]
\caption{List of the 8 \nustar\ PWN targets selected for this study}
\vspace{-0.7cm} 
\begin{center}
{\small 
\begin{tabular}{lcccccc}  
\hline\hline 
\noalign{\smallskip}
PWN name & TeV source name &  \edot\ [\lumcgs]  & Age [kyrs]$^a$ & Comments \\ 
\hline 
G18.5$-$0.4 (Eel) & HESS J1826$-$130 & $3.6\times 10^{36}$ & 14.4 & PeVatron candidate \\
G75.23$+$0.12 (Dragonfly) & eHWC J2019$+$368 & $3.6\times 10^{36}$ &  17.0 & PeVatron candidate  \\ 
G32.64$+$0.53 & eHWC J1850$+$001 & $9.8\times10^{36}$  & 42.9  & PeVatron candidate  \\ 
G106.65$+$2.96 (Boomerang) & VER J2227$+$608 & $2.2\times10^{37}$  & 10.4 & Leptonic or hadronic?$^b$ \\ 
G309.92$-$2.51 & HESS J1356$-$645 & $3.1 \times 10^{36}$ & 7.3 & Leptonic or hadronic?$^b$ \\  
G313.54$+$0.23  & HESS J1420$-$607  & $1.0 \times 10^{37}$ & 12.9 &  Crushed PWN \\
G313.3$+$0.1 (Rabbit) & HESS J1418$-$609 & $4.9 \times 10^{36}$ & 10.3  & Crushed PWN \\ 
G0.9$+$0.1 & HESS J1747$-$281 & $4.3 \times 10^{37}$ & 5.3 &  In the Galactic Center \\ 
 \hline 

\end{tabular} 

$^a$ Pulsar spin-down age. $^b$ It is unclear if the TeV source is of leptonic or hadronic origin. 
}
\vspace{-0.8cm}
\end{center}
\label{tab:nustar_targets}
\end{table}

\smallskip
\smallskip 
\noindent (1) \textbf{PeVatron PWN candidates (3 targets):}  G75.23$+$0.12 and G32.6$+$0.5 are associated with HAWC sources detected above $E_\gamma\sim50$~TeV, suggesting that they are PeVatron candidates. G18.5$-$0.4 (Eel PWN) is located in a complex region where several HAWC sources have been detected above 50 TeV (see \S\ref{sec:eel}). 

\smallskip 

\noindent (2) \textbf{Leptonic or hadronic TeV source? (2 targets):}  The X-ray morphology of G309.92$-$2.51 strongly supports the PWN scenario, but it is still unclear whether its TeV emission has a leptonic and/or hadronic origin. G106.65$+$2.96 (Boomerang PWN) may contribute to the TeV emission detected within SNR G106.3$+$2.7 which interacts with a molecular cloud  \citep{Amenomori2021} (see \S\ref{sec:boomerang}).  

\smallskip 

\noindent 
(3) \textbf{PWNe crushed by SNR reverse shocks (2 targets):}   G313.54$+$0.23 and G313.3$+$0.1 (the Rabbit nebula) are located in the complex Kookaburra region, which contains multiple high energy sources. Radio and X-ray morphology studies suggest that these extended X-ray nebulae offset from their TeV nebulae are potentially being crushed by SNR reverse shocks \citep{Aharonian2006}. 

\smallskip 

\noindent 
(4) \textbf{PWN in the Galactic Center:} 
G0.9$+$0.1 is a 5.3~kyr old PWN only $\sim1^\circ$ away from Sgr~A* \citep{Camilo2009}. This PWN is peculiar as its TeV emission is one of the weakest among the PWNe with similar spin-down ages. Despite some multi-wavelength SED studies, the nature of this unique PWN is uncertain with several possibilities: (1) the lack of low energy electron population due to fast adiabatic cooling \citep{Takata2011}, (2) lepto-hadronic scenario \citep{Holler2012} and (3)  highly-magnetized PWN \citep{vanRensburg2018}. 

\subsection{\nustar\ data analysis} 

We are currently analyzing \nustar\ spectral, imaging and timing data of the 8 PWN targets shown in Table~\ref{tab:nustar_targets}. We will jointly analyze available X-ray data from \chandra\ and \xmm\ observations in order to characterize the broad-band X-ray morphology/spectra of PWNe. We largely follow the \nustar\ data analysis of other PWNe such as DA 495 \citep{Coerver2019}. 
Upon detecting a pulsation signal, we can remove  photon events from the pulsed components using a folded lightcurve or extrapolate the pulsar's \chandra\ spectrum and estimate its contribution in the \nustar\ band -- these are the standard methods for mitigating the contamination of the pulsar \citep{An2019}. 
Since most of our targets are located outside the  Galactic ridge and thus do not suffer from the Galactic ridge X-ray emission background, the {\tt nuskybgd} tool \citep{Wik2014} will be used to model \nustar\ background images. This will help resolve subtle hard X-ray features in the  energy-resolved images by subtracting the simulated background images thus improving the signal-to-noise ratio  \citep{Nynka2014}. 

\subsection{\fermi-LAT data analysis} 

In between the X-ray and TeV bands, there are cases where \fermi-LAT GeV flux measurements or upper limits, together with other wavelength SED data, constrained the PWN parameters and demonstrated the unique role of GeV observations \citep{Principe2020}. Furthermore, the ICS photons in the GeV band, without fast synchrotron cooling, allow robust measurements of intrinsic electron spectra which reflect the particle acceleration mechanism in middle-aged PWNe.  To obtain GeV spectra of the PWNe, we are analyzing 13 years of LAT data using the latest Pass 8 version of the data with “Source” class events. By default, we use only events in the energy range from 10 GeV to 2 TeV due to the reduced contamination from diffuse Galactic emission and pulsar emission.  
To limit the contamination from gamma-ray emissions from the Earth’s limb, we set the maximum zenith angle set to $105^\circ$. The region of interest (ROI) is a square area of $10^\circ\times10^\circ$ centered at the position of each PWN. 
All point-like and extended sources within a square area of $30^\circ \times 30^\circ$ from the center of the ROI listed in LAT 10-year Source Catalog (4FGL-DR2) are included in the model. The Galactic diffuse emission and extragalactic isotropic emission are modeled using {\tt gll\_iem\_v07.fits}  
and {\tt iso\_P8R3\_SOURCE\_V3\_v1.txt}, respectively. We undertake the binned likelihood analysis with Fermipy ($0.1^\circ$ square bins), a python package that facilitates LAT data analysis with the \fermi\  Science Tools. 

\section{Preliminary results} 

We highlight our preliminary results on the Eel and Boomerang PWNe below. We plan to publish multiple papers, at least one from each source, including a comprehensive \nustar\ PWN paper summarizing the multi-wavelength observations.

\begin{figure*}%
    \centering
    (a){{\includegraphics[width=0.44\columnwidth]{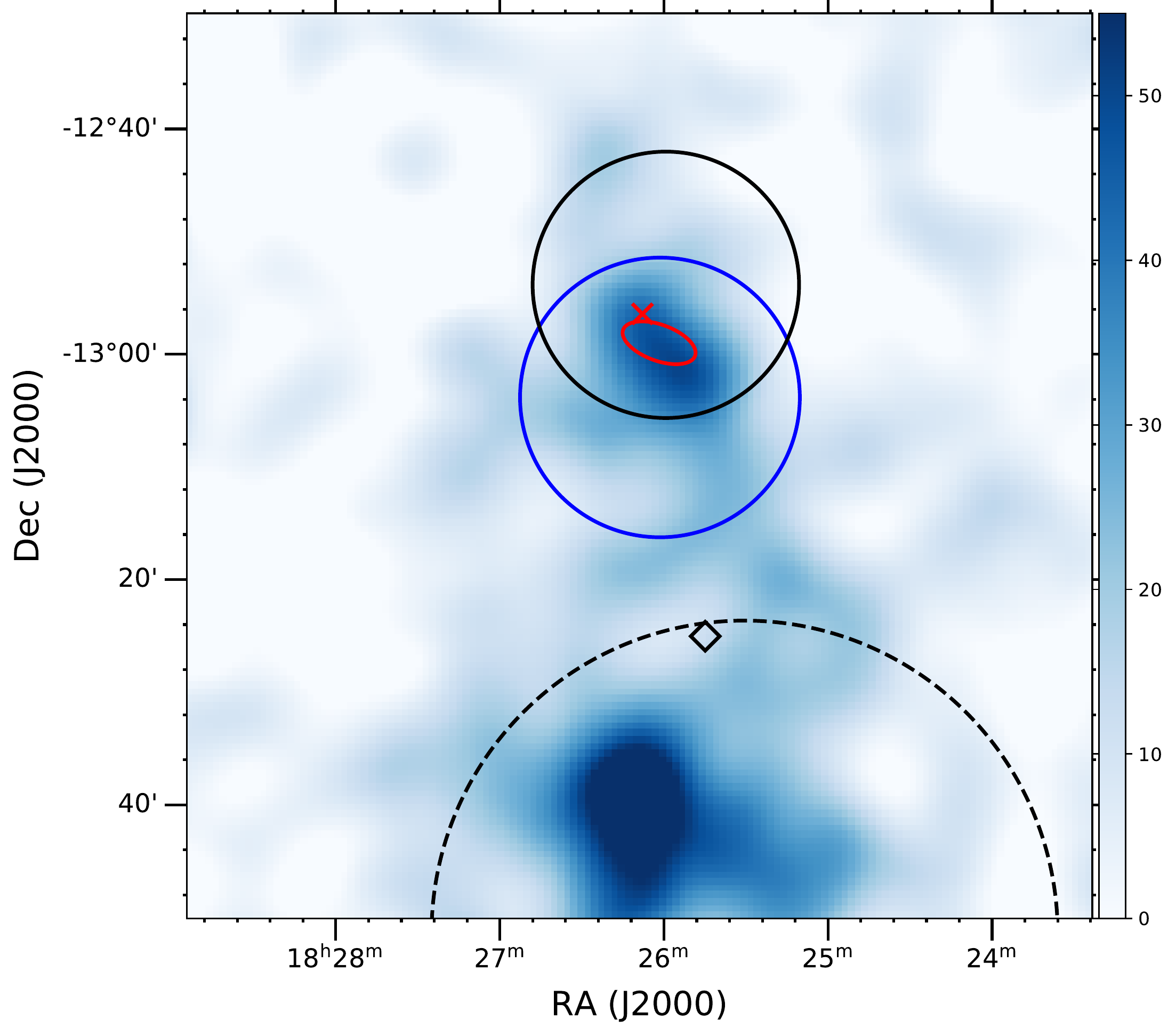}}}
    (b){\includegraphics[width=0.44\textwidth]{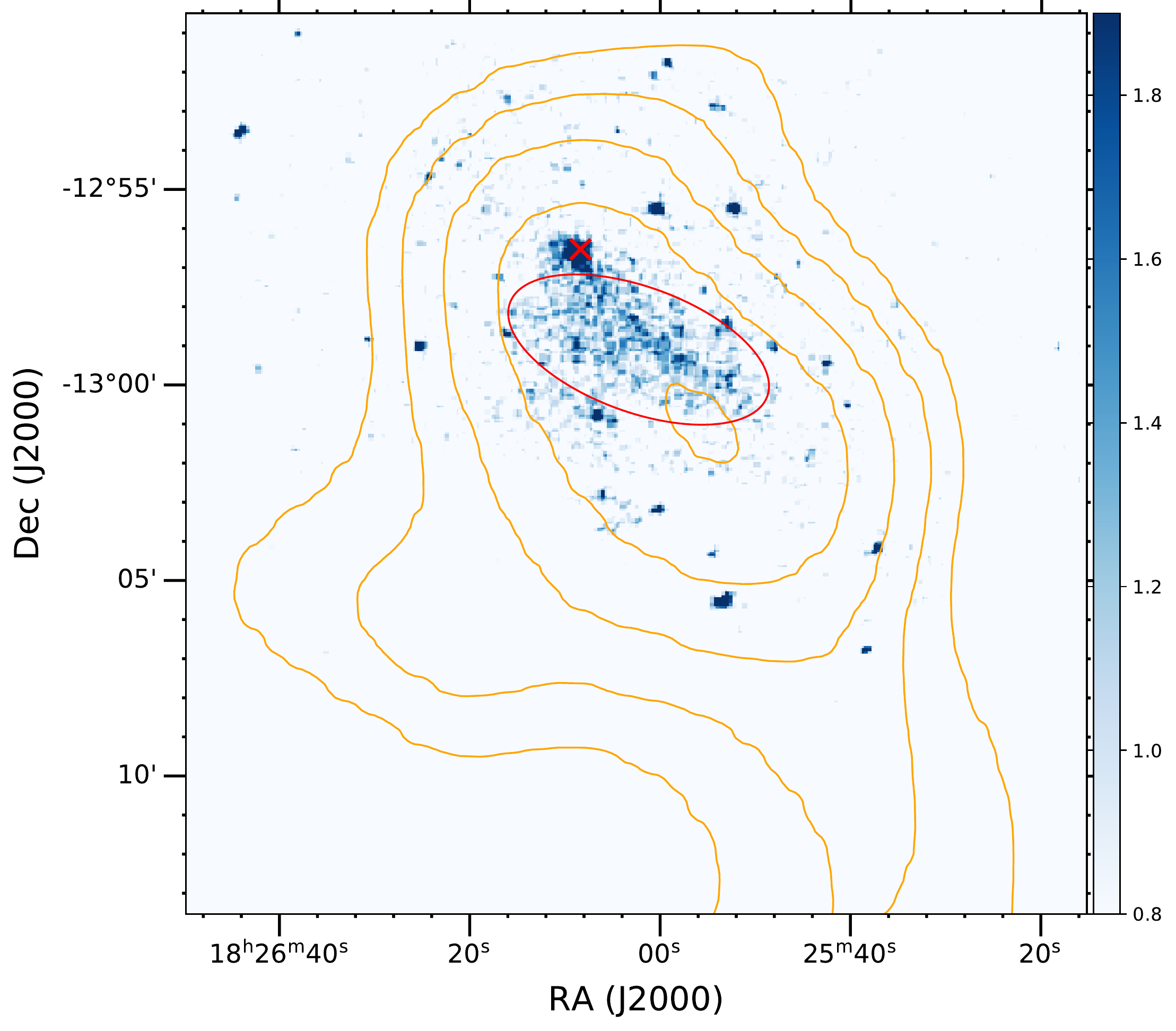}}%
    \caption{(a) H.E.S.S. $>10$ TeV excess map 
    of the region surrounding PSR J1826$-$1256. HAWC J1825$-$134 (solid black circle) and HESS J1826$-$130 (solid blue circle) are spatially coincident with the pulsar.     
    The other HAWC source's (HAWC J1825$-$138) extension to the south is shown with a dashed black circle.  
    (b) \xmm-MOS image of the Eel PWN. 
    H.E.S.S. source contours above 10 TeV are displayed in orange. In both figures, the pulsar's location is marked with a red cross and the red ellipse indicates the extended X-ray PWN. 
    }%
    \label{fig:eel_hess_xmm}%
\end{figure*}

\subsection{Eel PWN: X-ray counterpart of HAWC J1825$-$138 and HESS J1826$-$130} 
\label{sec:eel} 

\begin{figure*}%
    \centering
    (a) {{\includegraphics[width=0.40\textwidth]{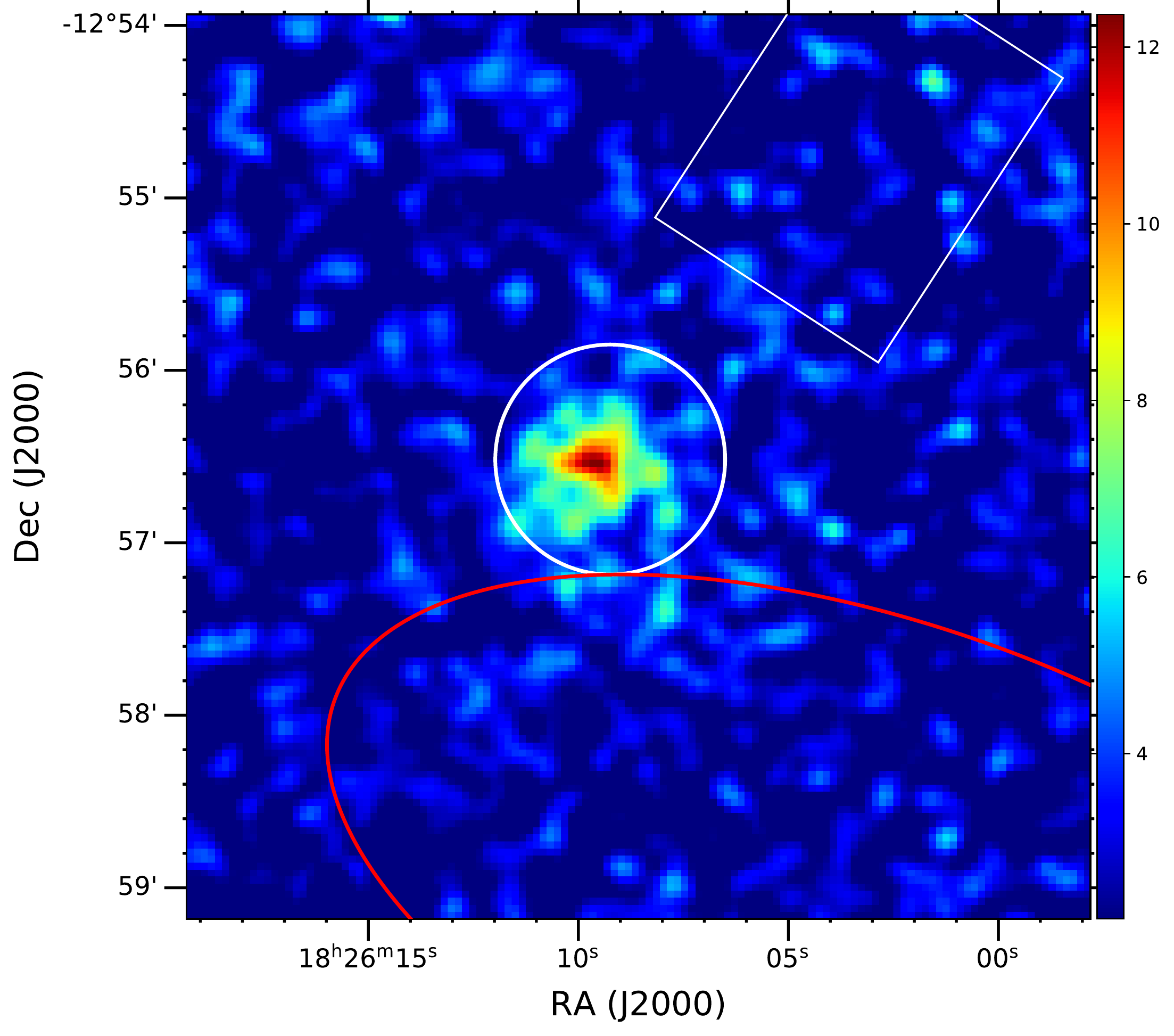} }} 
    (b) \includegraphics[width=0.5\columnwidth]{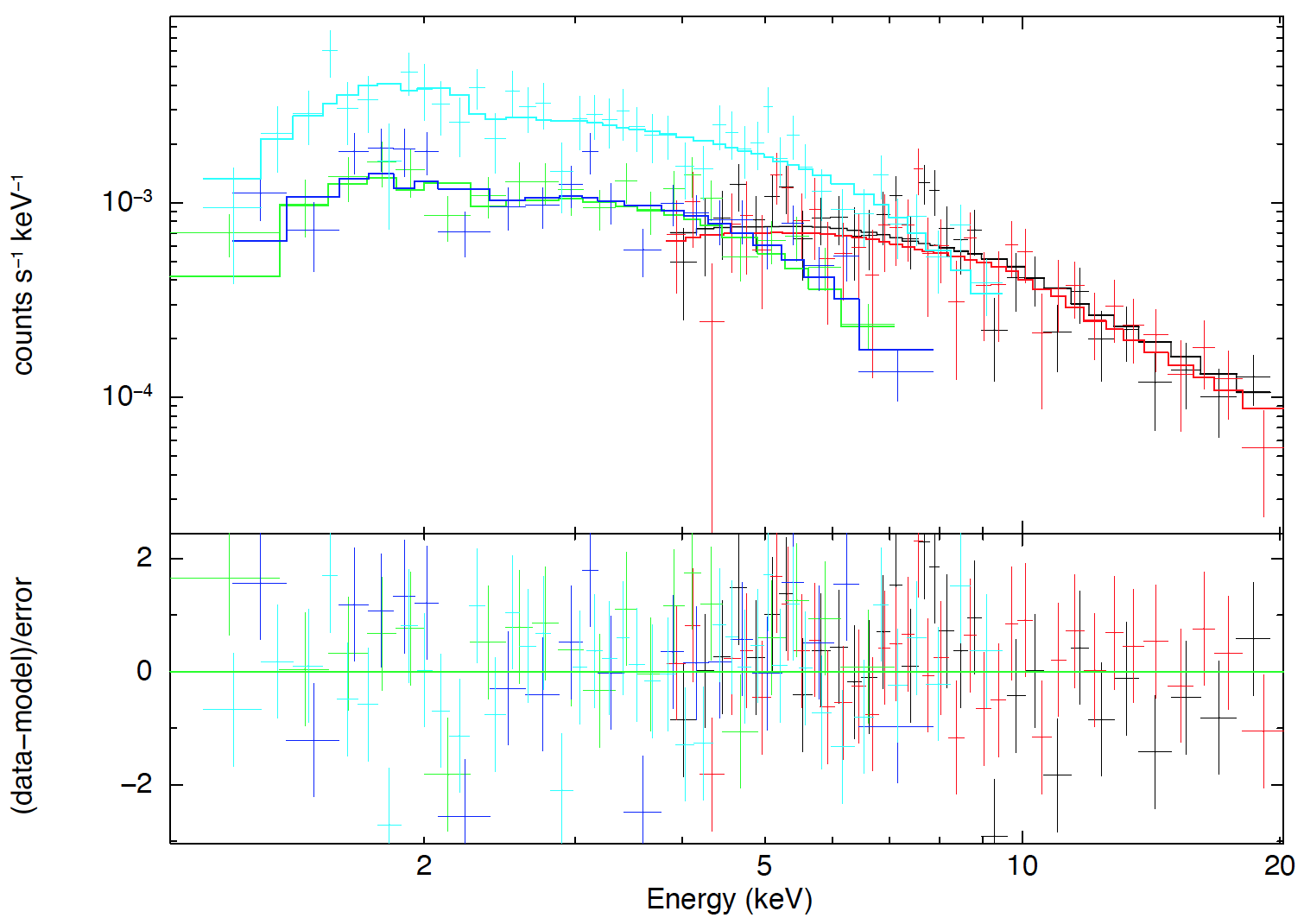} 
    \caption{(a) \nustar\ image of the Eel PWN in the 10--20 keV band.  
    Source and background extraction regions for \nustar\ spectral analysis are shown in the white circle  ($r=40$\asec) and white  rectangle, respectively. The red ellipse marks the extended PWN.  
    (b) Broad-band (1--20 keV) X-ray spectra of the inner PWN using \nustar\ FPMA (black) and FPMB (red) and \xmm\ EPIC MOS1 (green), MOS2 (blue) and PN (cyan). 
    The \nustar\ and \xmm\ spectra are jointly fit to an absorbed power-law model. }%
    \label{fig:eel_nustar}%
\end{figure*}

A GeV pulsar (PSR J1826$-$1256) and its extended X-ray PWN are roughly coincident with HAWC J1825$-$138 and HESS J1826$-$130.  
The pulsar has a characteristic age of 14.4 kyr, spin-down luminosity of  $3.6 \times 10^{36}$ erg s$^{-1}$, and surface magnetic field $B = 3.7 \times 10^{12}$ G \citep{Kargaltsev2017}. 
\xmm\ and \chandra\ X-ray observations have characterized  the surrounding PWN (known as the Eel PWN, \citep{Roberts2007}) in the X-ray band below 10 keV. The Eel PWN is extended, with a ``tail'' stretching over a  $3.5\amin\times1.6\amin$ ellipse in the X-ray band (Figure \ref{fig:eel_hess_xmm}; right panel). However, it also features more compact and intense X-ray emission within $r\sim15$\asec\  from the pulsar \citep{Karpova2019} (``inner PWN''), while the diffuse X-ray tail feature will be referred to as the ``extended PWN''. 
\nustar\ detected the inner PWN up to 20 keV (Figure \ref{fig:eel_nustar}) but failed to detect the extended PWN.  An absorbed power-law model fit to the \xmm\ and \nustar\ spectra of the inner nebula (Figure~\ref{fig:eel_nustar}; right panel) yields a photon index of $\Gamma = 1.32\pm0.08$.  
We found that a predominant fraction ($> 90$\% in the 2--10 keV band) of the X-ray emission originates from the extended X-ray nebula. Using \xmm-EPIC and \chandra\ ACIS data, we determine that the extended PWN has a softer photon index of $\Gamma = 2.03\pm0.15$ than that of the inner nebula.

\subsection{Boomerang PWN: a highly magnetized, crushed PWN?} 
\label{sec:boomerang} 

\begin{figure*}%
    \centering
    (a) {{\includegraphics[width=0.40\textwidth, angle=90]{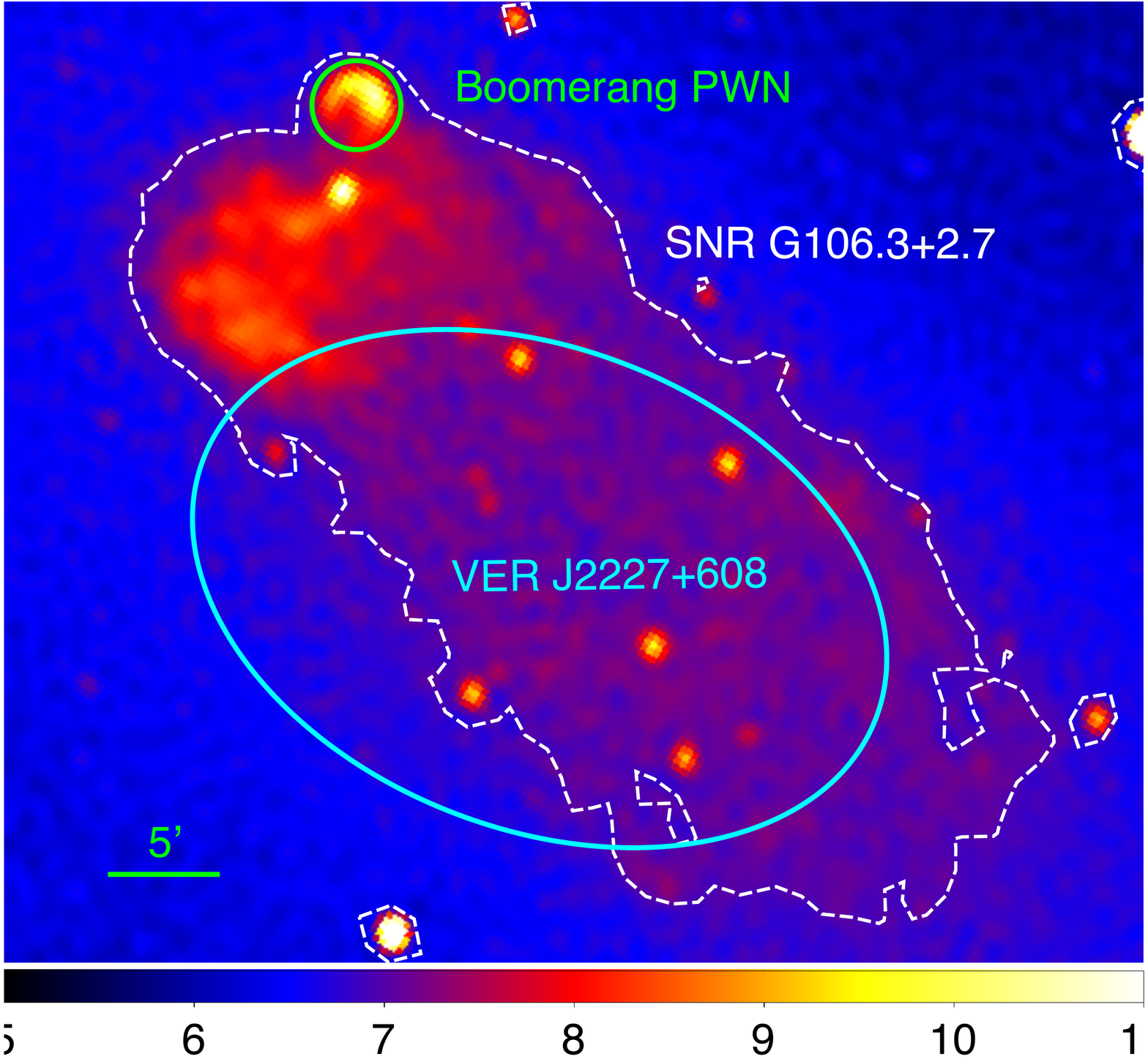} }} 
    (b) \includegraphics[width=0.37\textwidth]{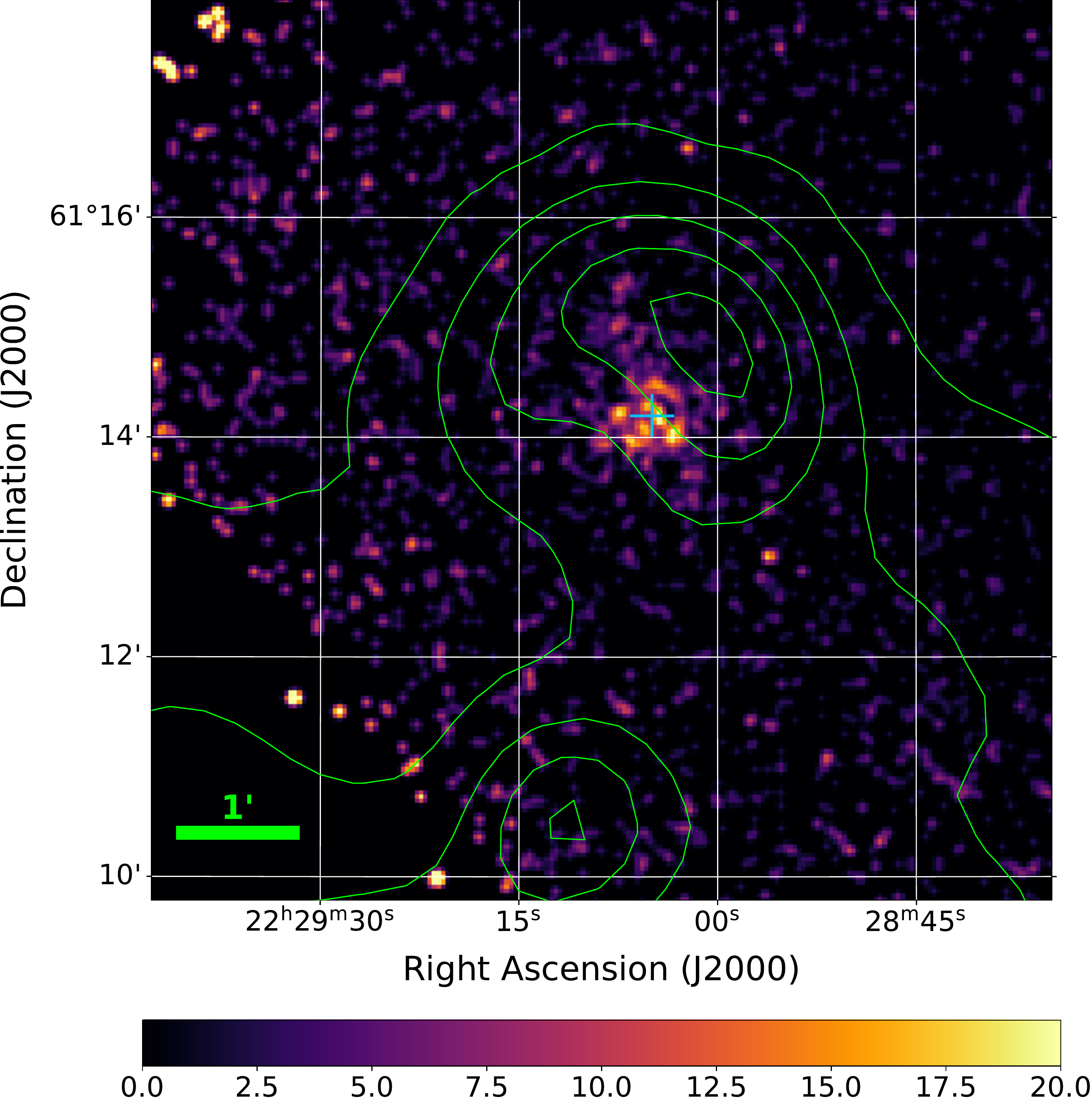} 
    \caption{(a) 1420 MHz radio image of the Boomerang region. The white dotted lines are radio contours indicating the boundary of SNR~G106.3$+$2.7. The X-ray extension of the PWN is marked by the green circle ($r=100$\asec) and the cyan ellipse represents the extension of TeV source VER~J2227$+$608.
    (b) \nustar\ 10--20 keV image of the Boomerang PWN. 1420 MHz radio contours are shown in green and the position of PSR~J2229$+$6114 is marked by the blue cross. 
    }%
    \label{fig:boomerang}%
\end{figure*}

The Boomerang region consists of a compact PWN around the radio pulsar PSR~J2229$+$6114 and its host SNR G106.3$+$2.7. The 51.6-ms pulsar has a characteristic age of 10.4 kyrs, spin-down luminosity of $2.2 \times 10^{37}$ erg s$^{-1}$ and surface magnetic field $B = 3.7 \times 10^{12}$ G, \citep{Halpern2001b}. 
The PWN's "boomerang" shape is  visible in the radio image of the region at 1420 MHz, and diffuse radio emission in the same band extends out opposite the apex of the PWN into its host SNR \citep{Kothes2001} (Figure~\ref{fig:boomerang}; left panel). 
The peculiar shape of the PWN and its extended radio emission suggests that the PWN has been crushed by a SNR reverse shock and the diffuse radio emission is a relic nebula swept away from the pulsar position \citep{Kothes2006,  Gaensler&Slane2006, Chevalier2011}. The SNR reverse shock scenario is further supported by a high PWN B-field ($B=2.6$ mG), which was suggested by a synchrotron break measured in the radio band \citep{Kothes2006}. 
In the GeV--TeV band, the Boomerang region has been extensively observed by \fermi-LAT and VERITAS. A recent detection by HAWC and LHAASO above $\sim50$~TeV suggests that this is a PeVatron candidate.   
The extended gamma-ray emission, detected by \fermi-LAT and VERITAS, is spatially coincident with the tail region of SNR~G106.3$+$2.7 as well as CO clouds \citep{Acciari2009}. In the X-ray band, both the pulsar and Boomerang PWN have been detected by \chandra\ \citep{Halpern2001b}. 
A recent study with \xmm, \chandra\ and \suzaku\ data detected diffuse X-ray emission from the head and tail region of the SNR  \citep{Ge2021}.

Our \nustar\ observation, pointing at the pulsar, detected extended X-ray emission from the Boomerang PWN up to 20 keV. We generate  background-subtracted \nustar\ images in the 3--10 keV and 10--20 keV bands (Figure~\ref{fig:boomerang}; right panel) and fit with a 2D Gaussian source profile and the \nustar\ PSF. We find the source size (FWHM) decreasing  from 35\asec\ (3--10 keV) to 20\asec\ (10--20 keV) toward higher energy, clear evidence of synchrotron burn-off in the PWN.  
Using the $Z^{2}$ algorithm, we searched for a pulsation from \nustar\ 3--20 keV lightcurve data extracted from a $r=30$\asec\ circular region around the pulsar. We detected a periodic signal at $P = 51.6$ msec, which is consistent with that measured by \citet{Halpern2001b}. We then produce  \nustar\ spectra of the PWN after removing photon events in the on-pulse phase interval. We fit the 3--20 keV \nustar\ PWN spectra to an absorbed power-law model with a photon index of $\Gamma = 1.65 \pm 0.08$.

\subsection{Multi-wavelength SED analysis} 

Our ultimate goal is to obtain the most accurate, broad-band SED data from radio to TeV band and determine fundamental PWN parameters by applying various SED models. First, we apply the standard \textsc{Naima} Python package for modeling SEDs  \citep{Zabalza2015}. 
In the \textsc{Naima} model, we assume  that the same underlying electron energy distribution, in the form of a power-law, cut-off power-law or broken power-law model, is producing synchrotron radiation in the radio to X-ray band and ICS  gamma-rays in the TeV band. Multi-wavelength SED fitting with \textsc{Naima} serves as our initial parameter survey and can be used to distinguish between the leptonic and hadronic models (in case that the origin of TeV emission is not firmly established). 
Our next step is to fit the SED data by a dynamical PWN evolution model \citep{Gelfand2009} which allows us to determine the PWN B-field, pulsar age and intrinsic electron energy distribution. 

\footnotesize 



%
%
%

\end{document}